\documentclass[prd,epsfig,twocolumn,floatfix,preprintnumbers,letterpaper]{revtex4}
\usepackage{graphicx}
\usepackage{amsmath,amssymb}
\usepackage{epsf}

\begin{document}

\pagestyle{myheadings}
\newcommand{\be}{\begin{equation}}
\newcommand{\ee}{\end{equation}}
\newcommand{\bea}{\begin{eqnarray}}
\newcommand{\eea}{\end{eqnarray}}

\title{Non-Minimal Quintessence With Nearly Flat Potential}
\author{Anjan A Sen{\footnote {email:anjan.ctp@jmi.ac.in}} and Gaveshna Gupta}
\affiliation{Center For Theoretical Physics,
Jamia Millia Islamia, New Delhi 110025, India}
\author{Sudipta Das}
\affiliation{Harish Chandra Research Institute, Chhatnag Raod, Jhunsi, 
 Allahabad 211019, India}
\begin{abstract}
We consider  Brans-Dicke type nonminimally coupled scalar field as a candidate for dark energy. In the conformally transformed Einstein's frame, our model is similar to {\it coupled quintessence} model. In such models, we consider potentials for the scalar field which satisfy the slow-roll conditions: $[(1/V)(dV/d\phi)]^2 << 1$ and $(1/V)(d^2V/d\phi^2) << 1$. For such potentials, we show that the equation of state for the scalar field can be described by a universal behaviour, provided the scalar field rolls only in the flat part of the potentials where the slow-roll conditions are satisfied. Our work generalizes the previous work by Scherrer and Sen \cite{scherrer} for minimally coupled scalar field case. We have also studied the observational constraints on the model parameters considering the Supernova and BAO observational data.

\end{abstract}
\maketitle

\section{INTRODUCTION :}
Over the last few years,there are growing evidences in favour of the scenario
that the universe at present is expanding with an acceleration. The supernovae 
type Ia observations \cite{obs1}, the cosmic microwave background radiation 
(CMBR) probes \cite{cmbr} suggest this acceleration very strongly. This is 
confirmed by the very recent WMAP data \cite{wmap} as well. These observations 
indicate that around 70\%
of the total energy density of the universe is in the form of an exotic,
negative pressure component dubbed as dark energy (See reference \cite{sami} 
for a recent review). The ratio of pressure to density for the dark energy 
component, referred to as the equation of
state parameter $\omega_{de}$, is given by
\be
\omega_{de} = \frac{p_{de}}{\rho_{de}}~.
\ee
Recent observations suggest that $\omega_{de}$ is close to $-1$. And 
assuming the equation of state parameter to be a constant, the approximate 
bound on $\omega_{de}$ is $-1.1 \le \omega_{de} \le-0.9$ (See references 
\cite{wood, davis}).
With future observational data, if $\omega_{de}$ continues to be around $-1$, then the most reasonable choice for 
dark energy will be a  cosmological constant. However, if the observations
predict a value close to $-1$, but not exactly equal to $-1$, dynamical dark energy models, like those which are driven
by a scalar field can be an interesting option to look for. Such models usually depict the evolution of $\omega_{de}$ with time.
This fact has recently been explored by Scherrer and Sen \cite{scherrer, 
anjan}, who examined the quintessence models as well as phantom dark energy 
models where
the scalar field potentials satisfy the slow-roll conditions :
\be\label{slowroll1}
\left({1\over{V}}{dV\over{d\phi}}\right)^2 << 1 
\ee
\be\label{slowroll2}
{1\over{V}}{{d^2}V\over{d{\phi^2}}} << 1.
\ee
The advantage of these models is that they provide a natural mechanism 
to produce an equation of state parameter $\omega_{de}$ slightly 
less than $-1$ at 
present. However the disadvantage of most of these scalar field models 
is that the quintessence potentials do not have any sound field 
theoretic background explaning their genesis. So, it might appear more 
appealing to employ a scalar field which is already there in the realm 
of the theory. This is where non-minimally coupled scalar field models 
become crucial where the scalar field is not put in by hand, but is 
already there in the purview of the theory. Although a long range scalar 
field as a gravitational field was first introduced by Jordan \cite{jordan}, 
this kind of non minimally coupled scalar field actually attracted the 
attention of the researcher when Brans and Dicke invoked such a field in 
order to incorporate the Mach's Principle in General Theory of 
Relativity \cite{bd}. Brans-Dicke (BD) theory, where the scalar field is 
directly coupled to the Ricci scalar, has the merit of producing results 
which can be compared with the corresponding GR results against 
observations \cite{obsbd}. A further important virtue of BD theory is that 
it is believed to produce the GR in a particular limit (see \cite{soma} 
for a different result). 

Possibility of having late time acceleration of the universe with 
nonminimally coupled scalar field ({\it a.k.a} scalar tensor theories) 
have also been explored in great details \cite{stquint}. Scaling 
attractor solutions with nonminimally coupled scalar fields have been 
studied with both exponential and power law potentials \cite{liddle}. 
Faraoni \cite{faraoni} has also studied with a nonminimal coupling 
term $(\phi^2/2)R$ with different scalar field potentials for the late 
time acceleration. Bertolo {\it et al.} \cite{bertolo}, Bertolami 
{\it et al.} \cite{orfeu}, Ritis {\it et al.} \cite{ritis} have found 
tracking soultions in scalar tensor theories with different type of 
potentials. In another work, Sen and Seshadri \cite{soma2} have obtained 
suitable scalar field potential to obtain power law acceleration of 
the universe. Saini {\it et al.} \cite{saini} and Boisseau {\it et al.} 
\cite{boss} have reconstructed the potential for a non minimally coupled 
scalar field from the luminosity-redshift relation available from the 
observations. Attempts have also been made to obtain an accelerating 
universe at present by introducing a coupling between the normal 
matter field and the Brans-Dicke field in a  generalised scalar tensor 
theory \cite{sudipta}.

\par In the present work we try to investigate dark energy models 
in non-minimally coupled theories of gravity, more precisely in BD theory. 
We write the equations in the
conformally transformed version of the theory (so called Einstein frame) 
and try to obtain a general form of the equation of state parameter 
$\omega_{de}$. 
The field equations in this
version look simpler and are more tractable. But now there is a 
coupling between the matter and the scalar field. This is in a way same 
as the so called {\it Coupled Quintessence} model extensively studied 
in the literature by number of authors \cite{coupled}. The existence of 
such interaction between the two components  can provide some ineteresting 
clue about the mystery of the dark energy \cite{mystery}. The effect of 
such coupling in different observational features like CMB and structure 
formation have also been studied \cite{coupobs}.
In the present work,we
try to obtain an expression for $\omega_{de}$ in the BD model in 
Einstein frame (which is essentially a coupled quintessence model) where 
the scalar field potential satisfies the slow roll conditions 
(\ref{slowroll1}) and (\ref{slowroll2}). 

\par The next section describes the field equations and their solutions. 
Section 3 describes the observational constraints on various
parameters of the model and the last section discusses the results. 

\section{FIELD EQUATIONS AND SOLUTIONS :}
The effective action for Brans-Dicke (BD) theory along with a self interacting
potential is given by
\be\label{action}
S=\int{\sqrt{-g}}~d^4x\left[\psi R - \frac{\omega}{\psi}
               {\psi}^{,\mu}{\psi}_{,\mu} -2 V(\psi) + L_{m}\right]~,
\ee 
where $R$ is the Ricci scalar, $\psi$ is the BD scalar field which is
nonminimally coupled to the gravity sector, $\omega$ is a dimensionless
parameter called the Brans-Dicke constant, $V(\psi)$ is the potential 
for the Brans-Dicke scalar field and $L_{m}$ represents the matter Lagrangian.
Here we have chosen the unit $8 \pi G = c = 1$.\\
For a spatially flat FRW universe, the line element is given by 
\be
ds^2 = dt^2 - a^2(t) \left[ dr^2 + r^2d{\theta}^2 + r^2\sin^2{\theta}
                      d{\phi}^2 \right]~.
\ee
Variation of action (\ref{action})
with respect to the metric components and the scalar field respectively yield
the Einstein field equations and the evolution equation for the scalar field as
\be\label{feq1}
 3\frac{\dot{a}^2}{a^2} = \frac{{\rho}_m }{\psi} + \frac{\omega}{2}
    \frac{\dot{\psi}^2}{\psi^2} - 3 \frac{\dot{a}}{a}\frac{\dot{\psi}}{\psi} +
    \frac{V}{\psi}~,
\ee
\be\label{feq2}
2\frac{\ddot{a}}{a} + \frac{\dot{a}^2}{a^2} = -\frac{\omega}{2}
      \frac{\dot{\psi}^2}{\psi^2} - \frac{\ddot{\psi}}{\psi}
              - 2\frac{\dot{a}}{a}\frac{\dot{\psi}}{\psi} +
	      \frac{V}{\psi}~, 
\ee
\be\label{waveeq}
\ddot{\psi} + 3 H \dot{\psi} = \frac{\rho_{m}}{2\omega +3} + \frac{1}{2\omega
  +3}\left[4V - 2\psi\frac{dV}{d\psi}\right]~.
\ee
From equations (\ref{feq1}), (\ref{feq2}) and (\ref{waveeq}) one can easily 
arrive at the energy conservation 
equation given by  
\be\label{matter}
\dot{\rho_{m}} + 3 \frac{\dot{a}}{a}\rho_{m} = 0~,
\ee
which is not an independent equation but follows from Bianchi identity.\\
Let us now effect a conformal transformation 
\be
\bar{g}_{\mu\nu} = e^{\frac{\phi}{\sqrt{\zeta}}}g_{\mu\nu}
\ee
where $\zeta = \frac{2\omega + 3}{2}$ and 
ln$\psi = \frac{\phi}{\sqrt{\zeta}}$. The conformal comoving time and the
scale factor are now related to the original ones through 
\begin{center}
$dt = e^{-\frac{\phi}{2\sqrt{\zeta}}}\bar{dt}~~~~~~$       and      $~~~~~~~ a =
  e^{-\frac{\phi}{2\sqrt{\zeta}}}\bar{a}$~.
\end{center}
The relevant field equations in the new frame look like 
\be\label{conf1}
3{\bar{H}}^2 = \bar{\rho}_{m} + \frac{1}{2}\dot{\phi}^2 + \bar{V}~,
\ee
\be\label{conf2}
2\dot{\bar{H}} + 3{\bar{H}}^2 = -\frac{1}{2}\dot{\phi}^2 + \bar{V}~,
\ee
\be\label{confwave}
\ddot{\phi} + 3 \bar{H} \dot{\phi} = \frac{\bar{\rho}_{m}}{2\sqrt{\zeta}} -
\frac{d\bar{V}}{d\phi}~,
\ee
where $\bar{\rho}_{m} = e^{-\frac{2 \phi}{\sqrt{\zeta}}}{\rho}_{m}$ and
$\bar{V} = e^{-\frac{2 \phi}{\sqrt{\zeta}}}V$. 
An overbar indicates quantities in the new frame and from now onwards an
overdot will indicate differentiation with respect to the transformed time
coordinate $\bar{t}$. The field equations in
the new frame are similar to that for coupled quintessence model previously studied \cite{coupled}.\\ 
In the new frame, the energy conservation equation also gets modified as 
\be\label{confmat}
{\dot{\bar{\rho}}}_{m} + 3 \bar{H} {\bar{\rho}}_{m} = -\sqrt{\frac{2}{3}}W
\dot{\phi}\bar{\rho}_{m}~, 
\ee
where $ W=\sqrt{\frac{3}{2}}\frac{1}{2\sqrt{\zeta}}$. Equation (\ref{confmat})
shows that the nonminimal coupling of the scalar field in the Brans-Dicke
theory leads to an explicit energy transfer between the
scalar field and the fluid in the Einstein frame.
Following references \cite{scherrer, holden}, equations
(\ref{conf1}) - (\ref{confwave}) can be 
written in the form of a plane-autonomous system by introducing the variables 
$x$, $y$ and $\lambda$ defined by 
\be
x=\frac{\dot{\phi}}{\sqrt{6}\bar{H}}~,
\ee
\be
y=\frac{\sqrt{\bar{V}}}{\sqrt{3}\bar{H}}~,
\ee
\be
\lambda=-\frac{1}{\bar{V}}\frac{d\bar{V}}{d\phi}~.
\ee
These give 
\be\label{omega}
{\Omega}_{\phi} = x^2 + y^2~,
\ee
and the equation of state parameter as 
\be\label{gamma}
\gamma \equiv 1 + \omega_{\phi} = \frac{2 x^2}{x^2 + y^2~}~.
\ee
Following the same method as ref.\cite{scherrer}, equations (\ref{conf1}) and
(\ref{confwave}) in a universe containing matter and scalar field (as we are
mainly interested in the late time behaviour of the universe, we neglect the
radiation component), become  
\be
x' = -3x + \lambda \sqrt{\frac{3}{2}}y^2 + \frac{3}{2}x (1 + x^2 - y^2) + W (1
- x^2 -y^2)~,
\ee
\be
y' = - \lambda \sqrt{\frac{3}{2}}x y + \frac{3}{2}y (1+ x^2 - y^2)~,
\ee
\be
\lambda' = -\sqrt{6} \lambda^2 (\Gamma - 1) x~,
\ee
where $\Gamma \equiv \bar{V}
\frac{d^2\bar{V}}{d\phi^2}/\left(\frac{d\bar{V}}{d\phi}\right)^2$ and a prime
indicates differentiation with respect to ln $a$. 
Since we are interested in models where $\omega_{\phi}$ is very close to
$-1$, it is convenient to express in terms of $\gamma$, which being very close
to zero, one can expand various quantities to lowest order in $\gamma$. 
Now using equations (\ref{omega}) and (\ref{gamma}), we change the dependent
variables $x$ and $y$ to the observable quantities ${\Omega}_{\phi}$ and
$\gamma$ and obtain 
\be\label{omegap}
{{\Omega}_{\phi}}' = 3 (1 - {\Omega}_{\phi})\left[(1 - \gamma)
  {\Omega}_{\phi} + \frac{W}{3}\sqrt{2 \gamma {\Omega}_{\phi}}
\right]~,
\ee
\bea\label{gammap}
\gamma' = &-&3 \gamma (2 - \gamma) + \lambda (2 - \gamma) \sqrt{3 \gamma
  {\Omega}_{\phi}}  \nonumber\\
&+&\sqrt{\frac{2 \gamma}{{\Omega}}_{\phi}} W (1 -
{\Omega}_{\phi}) (2 - \gamma)~,
\eea
\be\label{lambdap}
\lambda' = - \sqrt{3} \lambda^2 (\Gamma - 1) \sqrt{\gamma
  {\Omega}_{\phi}}~.
\ee
It is interesting to note that equations (\ref{omegap}), (\ref{gammap}) and
(\ref{lambdap}) are similar to those in ref \cite{scherrer}, 
except that now we have an
additional interaction term involving $W$ arising due to the non-minimal
coupling. 
If we switch off this interaction,
i.e, $W \rightarrow 0$, this indicates that the Brans-Dicke parameter $\omega
\rightarrow \infty$ and we get back the GR limit. This is consistent with the
notion that Brans-Dicke theory reduces to GR in the infinite $\omega$ limit.\\

From equations (\ref{omegap}) and (\ref{gammap}), one arrives at 
\bea\label{gammaomega} 
\frac{d\gamma}{d{\Omega}_{\phi}} &=& 
\frac{-3\gamma(2 - \gamma) + \lambda (2
- \gamma)\sqrt{3\gamma {\Omega}_{\phi}}} {3 (1 - \gamma) 
(1 - {\Omega}_{\phi}) \left[{\Omega}_{\phi} +
    \frac{W}{3}\sqrt{2 \gamma {\Omega}_{\phi}} (1 + \gamma) \right]}\nonumber\\
&+&
\frac{\sqrt{\frac{2\gamma}{{\Omega}_{\phi}}} W (1 - {\Omega}_{\phi}) (2 -
\gamma)}{3 (1 - \gamma) (1 - {\Omega}_{\phi}) \left[{\Omega}_{\phi} +
  \frac{W}{3}\sqrt{2 \gamma {\Omega}_{\phi}} (1 + \gamma) \right]}.\nonumber\\
\eea

\noindent
We should also emphasis that this equation is valid only when 
$d\Omega_{\phi}/dt \neq 0$ (See \cite{osc} for models where $\Omega_{\phi}$ 
oscillates in time). 

\noindent
At this point we make two assumptions - the first one being $\gamma << 1$,
which corresponds to $\omega_{\phi}$ very close to $-1$ as 
discussed earlier. The
second assumption we make is that the scalar field starts with an initial 
value in a potential which is nearly flat and the slow-roll 
conditions (\ref{slowroll1}), (\ref{slowroll2}) are satisfied.

\begin{figure}[!t]
\centerline{\epsfxsize=3.7truein\epsfbox{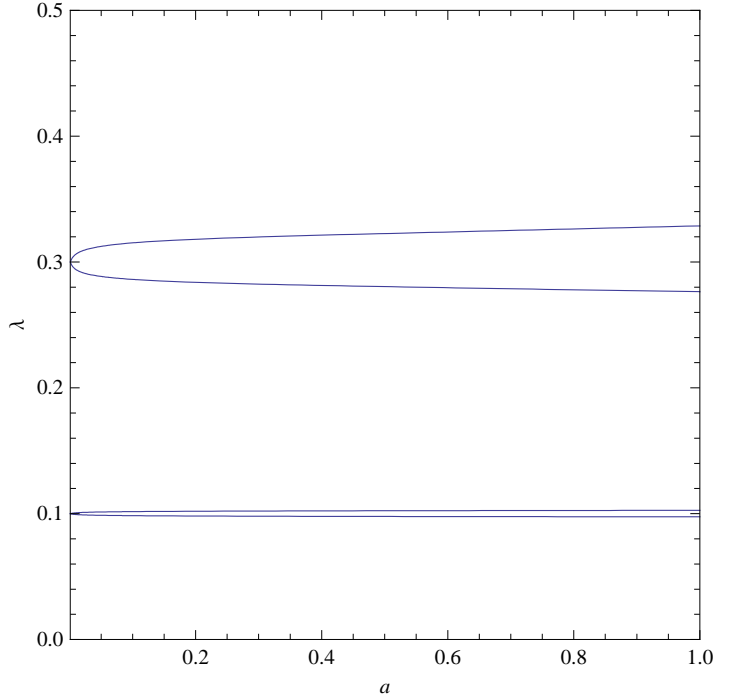}}
{\caption{\normalsize{Plot of $\lambda$ as a function of scale factor $a$. 
$W$ has been choosen to be 0.05. The lower set is for $\lambda_{0} =0.1$ 
and the upper set is for $\lambda_{0} = 0.3$. For each set the upper line 
is $V(\phi) = \phi^2$ and the lower line is for $V(\phi) = \phi^{-2}$.}}}  
\end{figure}

Assuming that these conditions are satisfied, one can use (\ref{lambdap}) 
to show that
$\lambda$ is approximately constant during the evolution of the 
scalar field for
different potentials, i.e $\lambda = \lambda_{0}$ where $\lambda_{0}$ 
is some initial
value corresponding to the inital scalar field value $\phi_{0}$. 
In figure 1, we have shown that this is indeed true. For a particular 
value of $W$, $W=0.05$, we have shown that the variation of $\lambda$ 
during the entire evolution is very small from its initial value 
$\lambda_{0}$ for two completely different potentials e.g 
$V(\phi) = \phi^2$ and $V(\phi) = {\phi^{-2}}$: for $\lambda_{0}=0.1$, 
$\lambda$ is practically constant, while for $\lambda_{0}=0.3$, 
the variation is ${\Delta \lambda\over{\lambda}} \leq 2.5\%$ from 
$a={10^{-3}}$ till today.
\begin{figure}[!t]
\centerline{\epsfxsize=3.7truein\epsfbox{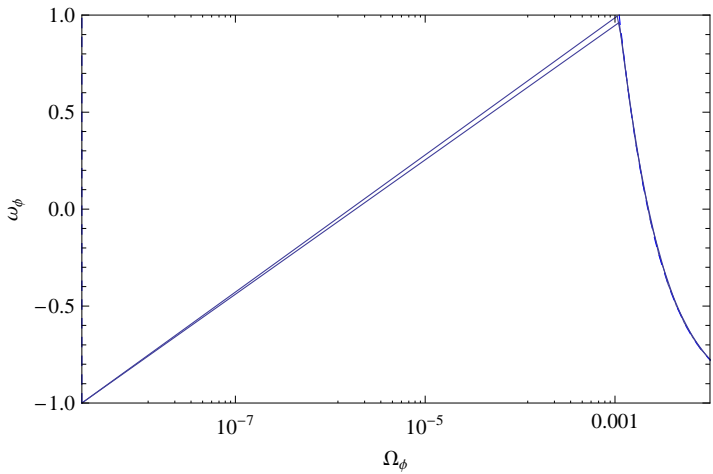}}
{\caption{\normalsize{Plot of $\omega_{\phi}$ vs. $\Omega_{\phi}$ for  
$W=0.05$ and $\lambda_{0} =0.1$. Different lines are for 
$V(\phi) = {\phi^2}$, $V(\phi) = {\phi^{-2}}$ and our analytical 
expression (\ref{eos}).They are all practically indistinguishable.}}}  
\end{figure}

So, we replace $\lambda$ by $\lambda_{0}$ in equation (\ref{gammaomega}) and
retain terms only to the lowest order in $\gamma$. Also we retain terms 
upto lowest order of $W$. Remember that $W$ is inversely related with 
the BD parameter $\omega$ which has to be large, resulting smaller value 
for $W$. This also ensures that the departure from the GR result is 
minimal. Using this, we obtain
following equation: 
\bea\label{dgamma}
\frac{d\gamma}{d{\Omega}_{\phi}} &=& \frac{-2\gamma}{\Omega_{\phi}
(1-\Omega_{\phi})} + \frac{2}{\sqrt{3}}\frac{\lambda_{0}\sqrt{\gamma}}
{\sqrt{\Omega_{\phi}}(1-\Omega_{\phi})}\nonumber\\
&+& \frac{2\sqrt{2}}{3}\frac{W\sqrt{\gamma}}{\Omega_{\phi}^{3/2}}
\eea

With $W =0$, one can get back to the solution $\gamma(\Omega_{\phi})$ 
as obtained earlier by Scherrer and Sen \cite{scherrer, anjan}. With $W$ 
one can also integrate the above equation to get 
the corresponding $\gamma(\Omega_{\phi})$. In order to do that one need 
to have some boundary condition to determine the integration constant 
arising due to integration of (\ref{dgamma}). The ideal condition should 
be $\gamma=0$ for $\Omega_{\phi} =0$. That is to say we start at very 
early time, when the dark energy contribution is zero and put the scalar 
field at the flat part of the potential. We can not take the limit 
$\Omega_{\phi}=0$ initially as term  involving $W$ in the expression 
for $\gamma$ after integration, blows up. Instead we assume that for 
some very small initial value of $\Omega_{\phi}$, say $\Omega_{i}$, 
$\gamma_{i} =0$. We should stress that $\Omega_{i}$ can be very very 
small but not exactly zero. With this initial condition, one can get 
the equation of state:
\bea\label{eos}
1+\omega_{\phi} =[\frac{\lambda_{0}}{\sqrt{3\Omega_{\phi}}} &-& 
\left(\frac{1}{\Omega_{\phi}}-1\right)[(\frac{\lambda_{0}}
{2\sqrt{3}}-\frac{\sqrt{2}W}{3})\nonumber\\
&\times&\log\left(\frac{1+\sqrt{\Omega_{\phi}}}{1-\sqrt{\Omega_{\phi}}}
\right) - \alpha]]^2,
\eea

\noindent
\begin{figure}[!t]
\centerline{\epsfxsize=3.2truein\epsfbox{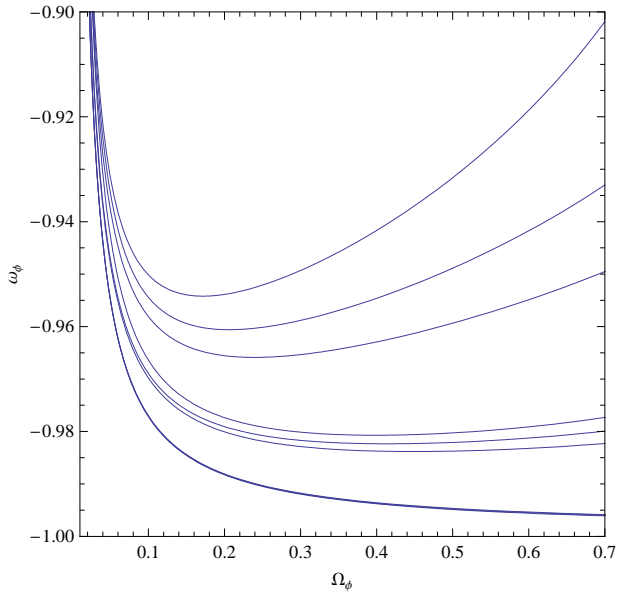}}
{\caption{\normalsize{Plot of $\omega_{\phi}$ vs. $\Omega_{\phi}$ for  
$W=0.05$. The three sets are for $\lambda_{0} =0.1, 0.2, 0.3$ from 
bottom to top. For each set, top line is for $V(\phi) = {\phi^2}$, the 
middle line is our analytical expression (\ref{eos}), and the bottom 
line is for $V(\phi) = {\phi^{-2}}$.}}}
\end{figure}

\noindent
where $\alpha = - \frac{\lambda_{0}}{\sqrt{3}}\frac{\sqrt{\Omega_{i}}
\Omega_{i}}{1-\Omega_{i}} - \frac{2\sqrt{2}W}{3}\sqrt{\Omega_{i}}$.  This 
is the main result we obtain in this study. This gives an analytical 
expression for the equation of state parameter for the dark energy as a 
function of its density parameter for a nonominimally coupled scalar 
field model (which is essentially a coupled quintessence model in 
Einstein frame). This generalizes the previous result obtained by 
Scherrer and Sen \cite{scherrer, anjan} for a corresponding minimally 
coupled scalar field model. The expression has an extra parameter $W$ 
which is related to the nonminimal coupling (or in other words related 
with the energy transfer between the matter and the dark energy in 
Einstein frame). The parameters $\lambda_{0}$ and $\Omega_{i}$ can be 
related with the present day value of the equation of state parameter 
$\omega_{\phi0}$ of the scalar field and the corresponding value of 
its density parameter $\Omega_{\phi0}$. Hence effectively the model has 
three parameters $W$, $\omega_{\phi0}$ and $\Omega_{\phi0}$. 
\begin{figure}[!t]
\centerline{\epsfxsize=3.2truein\epsfbox{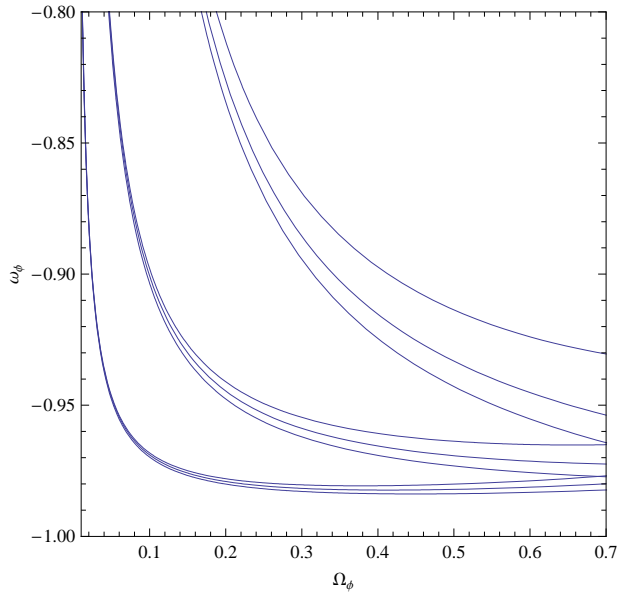}}
{\caption{\normalsize{Plot of $\omega_{\phi}$ vs. $\Omega_{\phi}$ for  
$\lambda_{0}=0.3$.The three sets are for $W = 0.05, 0.1, 0.2$ from bottom 
to top. For each set, top line is for $V(\phi) = {\phi^2}$, the middle 
line is our analytical expression (\ref{eos}), and the bottom line is 
for $V(\phi) = {\phi^{-2}}$.}}}  
\end{figure}
Next we want to show that our analytical expression (\ref{eos}) for the 
equation of state parameter $\omega_{\phi}$ indeed represents the true 
behaviour for $\omega_{\phi}$ for variety of potentials $V(\phi)$ subject 
to the condition that all of them have a very flat part over which the 
scalar field rolls, i.e, $\lambda_{0}$ is small. One peculiar feature of 
the nonminimal coupling (or coupled quintessence) is that even though we 
start with a scalar field which is initially confined to the flat part of 
the potential leading to $\omega_{\phi i} = -1$, it starts increasing to 
a higher positive value initially before coming back to a negative value 
near $-1$. And this is not only for our approximate analytical expressions 
for $\omega_{\phi}$ given in (\ref{eos}), it is also true when we solve 
numerically the exact equations (\ref{omegap}), (\ref{gammap}), and 
(\ref{lambdap}). And this is solely due to the interaction between the 
matter and dark energy. This interaction term is related with matter 
energy density (see equation (\ref{confwave})) and as initially it 
is large, the interaction is also very large. This actually violates 
our assumption of $\gamma << 1$ initially. But we should emphasis that 
this does not create any problem for the viable cosmology as this 
peculiar behaviour happens only when $\Omega_{\phi}$ has negligible 
contribution to the total energy density of the universe, and does not 
contribute in any way to the cosmological evolution. Once $\Omega_{\phi}$, 
starts contributing substantially to the total energy density, the 
equation of state $\omega_{\phi}$ reaches value close to cosmological 
constant which is what we expect. In Figure 2, we have shown the 
behaviour of $\omega_{\phi}$ as a function of $\Omega_{\phi}$ for 
negligibly small values of $\Omega_{\phi}$. We have shown the behaviours 
for analytical expression (\ref{eos}) as well as the result obtained by 
numerically solving (\ref{omegap})-(\ref{lambdap}) for potentials 
$V(\phi) = {\phi^2}$ and $V(\phi) = {\phi^{-2}}$. The plots are shown 
for $W=0.05$ and $\lambda_{0} = 0.1$ ($\Omega_{i}$ has been choosen so as 
to have $\Omega_{\phi}$ at present to be equal to 0.7). One can see the 
different behaviours are indistinguishable from each other, showing, 
(i) with small $\lambda_{0}$ different potentials have similar 
evolutions and (ii) our analytical expression (\ref{eos}) 
represents correctly the exact behaviour for small $\lambda_{0}$. 
\begin{figure}[!t]
\centerline{\epsfxsize=3.7truein\epsfbox{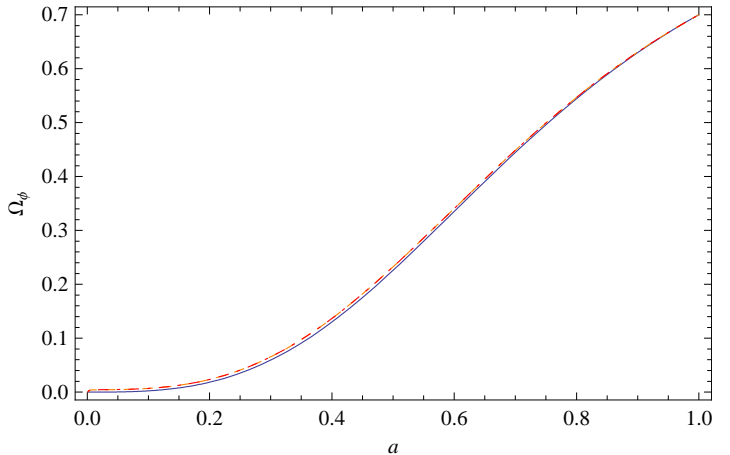}}
{\caption{\normalsize{Plot of $\omega_{\phi}$ vs. scale factor $a(t)$ 
for  $\lambda_{0}=0.3$, $\Omega_{\phi0} = 0.7$. The three sets are for 
$W = 0.1$. The solid line is for (\ref{omegaa}), the dashed line for 
$V(\phi) = {\phi^2}$, and the dotdashed line is for  $V(\phi) = {\phi^{-2}}$. 
As one can see all three lines are indistinguishable.}}}  
\end{figure}

In figure 3 and figure 4, we show the same behaviour 
$\omega_{\phi}(\Omega_{\phi})$ for larger values of $\Omega_{\phi}$. 
We have plotted $\omega_{\phi}$ as a function of $\Omega_{\phi}$ for 
our analytical result (\ref{eos}) as well as by solving numerically the 
exact equations (\ref{omegap}) - (\ref{lambdap}) for different potentials, 
using different values of $W$ and $\lambda_{0}$. In each case, $\Omega_{i}$ 
has been chosen in order to have $\Omega_{\phi}$ at present to be equal 
to 0.7. The figures show that for small $\lambda_{0}$ our analytical 
expressions nicely represent the exact behaviours for completely 
different potentials. 

Next we have to find the expression for $\Omega_{\phi}(a)$, in order to 
express $\omega_{\phi}$ as a function of the scale factor $a$. This is 
essential for confronting the model with any observational result. In this 
regard, we can use (\ref{omegap}) to solve $\Omega_{\phi} (a)$ to use it 
subsquently to determine $\omega_{\phi}(a)$. Assuming both $\gamma$ and 
$W$ to be small, we can neglect the terms involving those in 
(\ref{omegap}), to get an approximate solution for $\Omega_{\phi}$:

\be\label{omegaa}
\Omega_{\phi} = [1 + (\frac{1}{\Omega_{\phi0}} -1){a^{-3}}]^{-1}
\ee

\noindent
where $\Omega_{\phi0}$ is the the present day value of $\Omega_{\phi}$ and 
we normalize the present scale factor to $1$. This is same as what we get 
for a $\Lambda$CDM model which is not unexpected as we are considering 
scalar fields which have equation of state very close to $-1$. To show that 
this form for $\Omega_{\phi}$ very closely resembles with what one gets 
by exactly solving the (\ref{omegap})-(\ref{lambdap})  for small 
$\lambda_{0}$, we show the plots in figure 5. Here we plot our 
analytical expression (\ref{omegaa}) for $\Omega_{\phi}$ together with 
the result one obtains for $\Omega_{\phi}$ by solving numerically the 
exact equations (\ref{omegap})-(\ref{lambdap}). We have fixed $W=0.1$ 
and $\lambda_{0} = 0.3$ for this purpose and assume $\Omega_{\phi0} = 0.7$. 
One can see that all three lines are just indistinguishable, confirming 
the fact that our analytical approximation (\ref{omegaa}) for $\Omega_{\phi}$ 
is indeed robust. It also shows the contribution of the scalar field to 
the total energy density of the universe is negligibly small till 
$a \sim 0.2$ and any peculiar behaviour of the equation of state 
$\omega_{\phi}$ before this era does not matter in any way.

So (\ref{eos}) together with the (\ref{omegaa}) for $\Omega_{\phi}(a)$ 
represents an analytical behaviour for the equation of state of a 
non-minimally coupled scalar field with BD type coupling. This behaviour 
is only valid when the scalar field rolls over  
a very flat part of the potential so that equations  
(\ref{slowroll1}) and (\ref{slowroll2}) are satisfied, irrespective of 
the form of the potential. This is main result of this study and 
its generalizes the similar result obtained earlier by 
Scherrer and Sen \cite{scherrer, anjan} for minimally coupled scalar fields.

\section{Observational Constraints}
\begin{figure}[!t]
\centerline{\epsfxsize=3.7truein\epsfbox{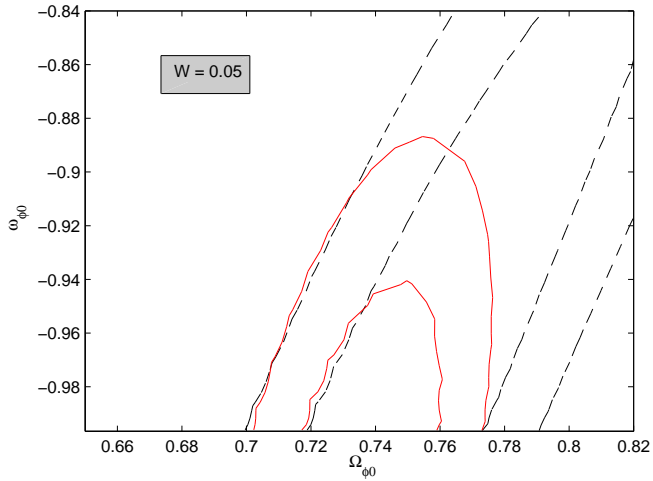}}
{\caption{\normalsize{$1\sigma$ and $2\sigma$ confidence contours in 
$\omega_{\phi0}-\Omega_{\phi0}$ plane. The dashed lines are for SnIa data 
only and the solid lines are for SnIa+Bao data (explained in the text). 
This is for $W=0.05$. }}}  
\end{figure}

Next we compare our result for $\omega_{\phi}(a)$ to the SnIa and 
Baryon Acoustic Oscillations data. The confidence contours are 
constructed using 60 Essence supernovae, 57 SNLS (Supernova Legacy Survey) 
and 45 nearby supernoave, and the data released by 30 SnIa detected by 
HST and classified as Gold sample by Riess et al \cite{riess}. The combined 
data set can be found in Ref. \cite{davis}. We also add the result 
obtained by Sloan Digital Sky Survey regarding the Baryon Acoustic 
Oscillation (BAO) Peak, giving a distance scale at $z_{bao} = 0.35$. 
In Figure 6,7 and 8 we have shown the confidence contours for $W=0.05$, 
$W=0.1$ and $W=0.2$.
\begin{figure}[!t]
\centerline{\epsfxsize=3.7truein\epsfbox{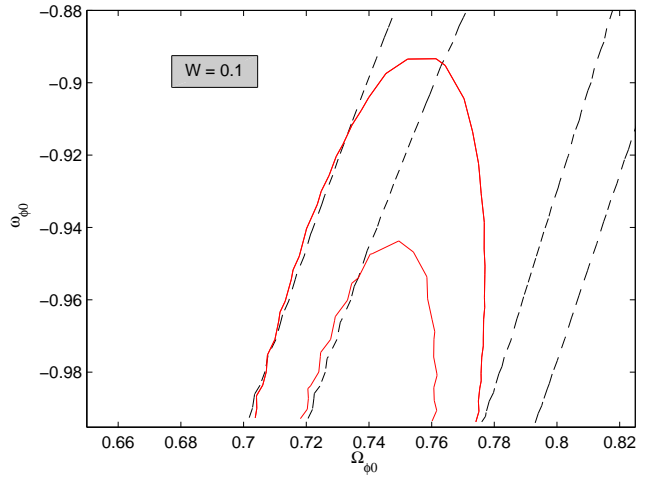}}
{\caption{\normalsize{$1\sigma$ and $2\sigma$ confidence contours in 
$\omega_{\phi0}-\Omega_{\phi0}$ plane. The dashed lines are for SnIa data 
only and the solid lines are for SnIa+Bao data (explained in the text). 
This is for $W=0.1$. }}}  
\end{figure}

\begin{figure}[!t]
\centerline{\epsfxsize=3.7truein\epsfbox{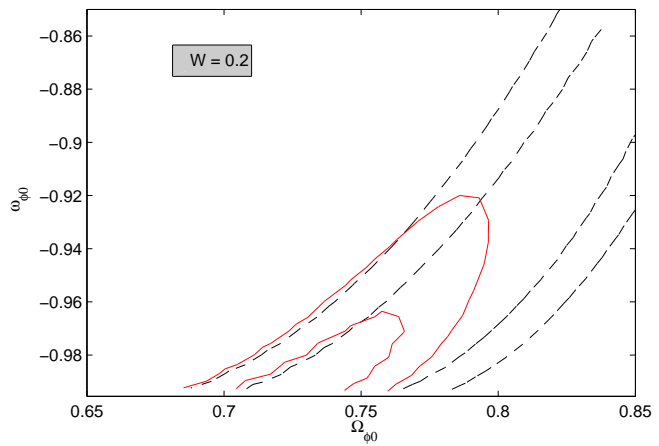}}
{\caption{\normalsize{$1\sigma$ and $2\sigma$ confidence contours in 
$\omega_{\phi0}-\Omega_{\phi0}$ plane. The dashed lines are for SnIa data 
only and the solid lines are for SnIa+Bao data (explained in the text). 
This is for $W=0.2$. }}}  
\end{figure}

Our result shows that cosmological constant is obviously very much 
consistent with the observational result. Also, unless one obtains a 
stringent bound on $\Omega_{m0}$ from structure formation data e.g the 
shape parameter for the matter power spectrum, which will result a 
corresponding bound on $\Omega_{\phi0}$, one can not put a strong 
constraint on $\omega_{\phi0}$, although BAO data seems to provide a 
strong upper bound for $\omega_{\phi}$ which is around 
$\omega_{\phi0} \leq -0.92$.

\section{Conclusion}
We have studied dark energy models with scalar fields which are 
non-minimally coupled to gravity. We assume the form of the coupling is 
of BD type. We write the system of equations in conformally transformed 
Einstein frame where the scalar field becomes coupled with the matter 
sector although minimally coupled with the gravity sector. Our model is 
exactly similar to the coupled quintessence models considered previously 
by several authors \cite{coupled}. To study such coupled quintessence 
models, we have assumed that the coupling parameter $W$ between the 
matter and dark energy (which is related with BD parameter $\omega$ 
in Jordan frame) is small. This ensures that our model does not deviate 
much from the standard Einstein gravity.

With a such a system, we consider potentials for the scalar fields 
which satisfy the slow-roll conditions (\ref{slowroll1}), (\ref{slowroll2}). 
In such  a scenario, we show that all of them converge to a universal 
behaviour given by (\ref{eos}) and (\ref{omegaa}). This work generalizes 
the previous result by Scherrer and Sen obtained for minimally 
coupled scalar fields, to the non-minimally coupled case which can also 
be described as a coupled quintessence model in Einstein frame. 

In this model, one has to fine tune the initial conditions in order to 
get desired value for the equation of state or density parameters at 
present which is in contrary to the so-called tracker model. But here 
the subsequent evolution of the universe is insensitive to the form of 
the potential as long as the slow-roll conditions (\ref{slowroll1}), 
(\ref{slowroll2}) are met. This is in contrast to the tracker models 
where one has to really fine tune the shape of the potential in order to 
achieve the tracker behaviour.

We have also tested our model with the observational data from SnIa 
and BAO peak. Although to get strong bound on the equation of state, one 
needs to have strong independent observational constraint on 
$\Omega_{\phi0}$, still inclusion of BAO data put a strong upper bound 
on $\omega_{\phi0}$ irrespective of the value of $W$, the coupling parameter.

\section{Acknowledgement :}
AAS acknowledges the financial support provided by the University 
Grants Commission, Govt. Of India, through the major research project 
grant (Grant No:  33-28/2007(SR)). Part of the work has been done at 
IUCAA, Pune (India) during the visit of AAS under associateship program.
SD would like to thank Relativity and Cosmology Research Centre, 
Jadavpur University, Kolkata where part of this work has been done.

\end{document}